\documentstyle[12pt]{article}
\input epsf.tex

\def \beq { \begin{equation} }
\def \eeq { \end{equation} }
\def \dum {\partial ^{{}^{\!\!-\!1}}\!\!\!}
\def \0j {j_{{}_0} }
\def \1j {j_{{}_1} }

\begin{document}

\title{Algebra of Non-Local Charges in Supersymmetric Non-Linear Sigma
Models}

\author{L.E. Saltini\thanks{Work supported by FAPESP. E-mail:
lsaltini@uspif.if.usp.br}$\;$ and A. Zadra\thanks{Work partially supported by
CNPq. E-mail: azadra@uspif.if.usp.br}\\
Instituto de F\'\i sica da Universidade de S\~ao Paulo\\
C.P. 66318, S\~ao Paulo, SP, 05389-970, Brazil}

\date{}

\maketitle

\begin{abstract}
We propose a graphic method to derive the classical algebra (Dirac brackets)
of non-local conserved charges in the two-di\-men\-sio\-nal
supersymmetric non-linear
$O(N)$ sigma model. As in the purely bosonic theory we find a cubic Yangian
algebra. We also consider the extension of graphic methods to other integrable
theories.
\end{abstract}

\section{ Introduction}

Non-linear sigma models [1-3] are prototypes of a remarkable class of
integrable two dimensional models which contain an infinite number of
conserved local and non-local charges [4-7].
The algebraic relations obeyed by such charges are supposed to be
an important ingredient in the complete solution of
those models [8-11]. The local charges form an Abelian algebra. Opposing to
that simplicity, the algebra of non-local charges is non-Abelian and
actually non-linear [12-28].

In ref.[29] the $O(N)$ sigma model was investigated and a particular set
of non-local charges -- called {\it improved} charges -- was found to satisfy
a cubic algebra related to a Yangian structure.
In this work we intend to extend that result to the corresponding
supersymmetric case [30-32]. The introduction of supersymmetry
might have rendered a much more involved algebra [33]. However, it has been
conjectured [29,32] that, in the sigma model, the algebra of supersymmetric
non-local charges would remain the same as in the bosonic theory and we shall
present results that confirm such conjecture.

This paper is organized as follows. In Sect.2 we briefly review the results
from the purely bosonic theory. A graphic technique to compute charges and
their algebra is introduced in Sect.3. In Sect.4 we discuss the supersymmetric
model and the main results of the paper. Another application of graphic rules
is shown in Sect. 5 concerning the $O(N)$ Gross-Neveu model. Sect.6 is left
for conclusions while an appendix contains examples of the graphic technique.

\section{ Bosonic model - a review}

The two-dimensional non-linear $O(N)$ sigma model can be described by the
constrained Lagrangean
\beq
{\cal L} = {1\over 2}\partial _\mu \phi_i\partial^\mu \phi_i \quad ,\quad
\sum _{i=1}^N \phi^2_i =1 \quad .
\eeq
Associated to the $O(N)$ symmetry we have a matrix-valued conserved
curvature-free current
\begin{eqnarray}
& &(j_\mu)_{ij}=\phi _i\partial _\mu \phi _j - \phi _j \partial _\mu \phi _i
\qquad ,\qquad \partial _\mu j^\mu = 0\quad ,\nonumber \\
& &f_{\mu \nu }=
\partial _\mu j_\nu - \partial _\nu j_\mu + 2[j_\mu ,j_\nu ] = 0
\qquad ,
\end{eqnarray}
whose components satisfy the algebra [29]
\begin{eqnarray}
& &\{ (\0j )_{ij}(x) , (\0j )_{kl}(y) \} = (I \circ  \0j )_{ij,kl}(x)
\delta(x-y) \nonumber \\
& &\{ (\1j )_{ij}(x) , (\0j )_{kl}(y) \} = (I \circ   \1j )_{ij,kl}(x)
\delta(x-y) + (I \circ   j)_{ij,kl}(x) \delta'(x-y) \\
& &\{ (\1j )_{ij}(x) , (\1j )_{kl}(y) \} = 0 \nonumber \quad .
\end{eqnarray}
where $I$ is the $N\times N$ identity matrix.
Above we have introduced the intertwiner field
\beq
(j)_{ij} = \phi_i\phi_j
\eeq
and the $O(N)$ $\circ $-product defined in ref. [29] as
\beq
(A\circ B)_{ij,kl}\equiv A_{ik}B_{jl} - A_{il}B_{jk} +
A_{jl}B_{ik} - A_{jk}B_{il}\quad .
\eeq
This model is known to have infinite non-local conserved charges. The standard
set of charges can be iteratively built up by means of the potential
method of Br\'ezin {\it et. al.} [5]. However, in
ref. [29] an alternative set of {\it improved} charges $\{ Q^{(n)} ,
n=0,1,2,\cdots \}$ has been defined and it was shown that they obey the
non-linear algebra
\beq
\{ Q^{(m)}_{ij}, Q^{(n)}_{kl}\} = \left( I\circ   Q^{(n+m)}\right)_
{ij,kl} - \sum _{p=0}^{m-1}\sum _{q=0}^{n-1}
\left( Q^{(p)} Q^{(q)} \circ Q^{(m+n-p-q-2)}\right)_{ij,kl} \quad .
\eeq
These charges were named {\it improved} because they brought up an algebraic
improvement: the non-linear part of the algebra is simply cubic, as opposed to
the algebra of the standard charges previously used in the literature [14]. The
Jacobi identity and other properties of the improved cubic algebra were
thoroughly discussed in ref. [29]. But there is a way to abbreviate that
algebra, which is the first among the new results of this paper and which will
be presented now.

\vskip .5truecm
We shall define a Hermitean generator of improved charges
\beq
Q(\lambda ) \equiv I + i\sum _{n=0}^\infty \lambda ^{n+1} Q^{(n)} \quad ,
\eeq
where $\lambda $ will be called the spectral parameter. Therefore one can
summarize the algebra (6) as follows:
\newpage
\beq
i\{ Q_{ij}(\lambda ),Q_{kl}(\mu )\} = \left( f(\lambda, \mu )\circ Q(\lambda )-
Q(\mu )\right) _{ij,kl}\quad ,
\eeq
where
\beq
f(\lambda ,\mu )\equiv \Re e\left( {Q(\lambda )Q(\mu )\over \lambda ^{-1} -\mu
^{-1}}\right) = {I - \sum _{m,n=0}^\infty \lambda ^{m+1}\mu
^{n+1}Q^{(m)}Q^{(n)}\over \lambda ^{-1} -\mu ^{-1}}\qquad .
\eeq
The quadratic non-linearity encoded in $f(\lambda ,\mu )$
can be related to the known Yangian structure that underlies this model
[17-26,29].
The advantage in writing the algebra as in (8) is not only aesthetic.
Recalling the monodromy matrix of standard charges, and its algebra expressed
in terms of the classical $r$-matrix,
\begin{eqnarray}
& &T(\lambda )={\rm exp} \sum _{n\ge 0}\lambda ^{n+1} \hat Q^{(n)} \quad
,\nonumber \\
& & \{ T(\lambda ){\buildrel \otimes \over ,} T(\mu ) \} =[r(\lambda ,\mu ),
T(\lambda )\otimes T(\mu )] \quad , \\
& &r(\lambda ,\mu )={I_a\otimes I_a\over \lambda ^{-1}-\mu ^{-1}}\quad
, \quad [I_a, I_b] = f_{abc}I_c \quad ,\nonumber
\end{eqnarray}
we remark that the generator $Q(\lambda )$ and the $f$-matrix play similar
r\^oles to those of the monodromy matrix and classical $r$-matrix in the
standard approach [17-26]. We do not fully understand the relationship between
(8) and (10) but we expect to be able to use this analogy to
establish a precise translation between the different sets of charges
[35]. We also hope that a complete knowledge about the conserved charges and
their algebra will become an decisive ingredient in off-shell scattering
calculations.

\vskip .5truecm
Now let us consider the graphic methods announced in the Introduction. We
recall that in ref. [29] the improved charges
were constructed by means of an iterative
algebraic algorithm that uses $Q^{(1)}$ as a step-generator, as indicated by
the relation
\beq
(I\circ Q^{(n+1)}) = {\rm linear\, part\, of\,} \{Q^{(1)},Q^{(n)}\} \qquad .
\eeq
``After a tedious calculation" the authors in ref. [29] managed to construct
the charges $Q^{(n)}$ and their algebra up to $n=5$. In the next section we
will present a {\it graphic} method that makes the calculation simpler, less
tedious and convenient for a further supersymmetric extension.

\section{ Graphic rules for the bosonic model}

Let us associate white and black semicircles to the $O(N)$ current components,
\beq
\0j \Longleftrightarrow \; \raise -.1cm \hbox{\epsfbox{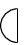}}
\hskip 2truecm
\1j \Longleftrightarrow \; \raise -.1cm \hbox{\epsfbox{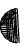}}
\eeq
a continuous line and an oriented line to the identity and the anti-derivative
operator respectively,
\beq
I \Longleftrightarrow \; \raise .07cm \hbox{\epsfbox{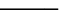}} \hskip 2truecm
2\dum \Longleftrightarrow \; \raise .07cm \hbox{\epsfbox{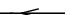}}
\eeq
The operator $\dum $ above follows the same convention adopted in ref. [29],
\beq
\dum A(x) ={1\over 2}\int {\rm d} y \, \epsilon(x-y)A(y)
\quad ,\quad \epsilon(x)= \cases{-1,\,& $x<0$\cr \phantom{-}0,\, &$x=0$\cr
+1,\, &$x>0$\cr}\quad .
\eeq
Below one finds some diagrams and the corresponding expressions:
\begin{eqnarray}
& &2\0j \dum \0j \qquad \Longleftrightarrow \qquad \raise -.1cm \hbox{
\epsfbox{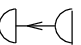}} \nonumber \\
& &2\dum \1j \0j \qquad \Longleftrightarrow \qquad \raise -.1cm \hbox{
\epsfbox{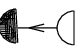}} \nonumber \\
& &4\dum \0j \dum \0j \qquad \Longleftrightarrow \qquad \raise -.1cm \hbox
{\epsfbox{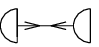}} \nonumber \\
& &4\1j \dum (\0j \dum \0j )\qquad \Longleftrightarrow \qquad \raise -.1cm
\hbox{\epsfbox{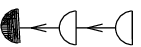}}
\end{eqnarray}
We have noticed [29] that every improved charge can be written as an integral
over symmetrized chains of $j_o$'s and $j_{{}_1}$'s connected by the operator
$2\partial ^{-1}$. Therefore we can associate a diagram to each improved
charge, as exemplified by the second non-local charge $Q^{(2)}$:
\beq
Q^{(2)}=\int dx \left[
2\0j + \0j (2\dum \1j ) + \1j (2\dum \0j ) + \0j (2\dum (\0j 2\dum \0j ))
\right]
\eeq
\[
\hskip 1.7cm \epsfbox{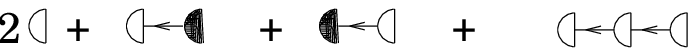}
\]

\noindent
If one is interested in constructing charges, there is an iterative graphic
procedure, inspired by the lessons taken from ref. [29] and which
will be described now. Consider the following properties:
\vskip .3truecm

(a) The improved non-local charges have the general form $Q^{(n)} = \int dx\,
J^{(n)}$ where $J^{(n)}$ is a combination of terms which one can always write
as
\beq
j_\mu \qquad {\rm or} \qquad 2(j_{\mu}\dum S+\dum S^{{}^t}j_{\mu}) \quad ,
\eeq
where $S$ is some chain and $S^t$ its transposed.

\vskip .5truecm
(b) The algebraic definition of improved charges is
\beq
(I\circ Q^{(n+1)})_{ij,kl}= {\rm linear \,\,\, part \,\,\, of \,\,\,}
\{Q_{ij}^{(1)},Q_{kl}^{(n)}\}
\eeq
and we note that the linear part of $\{ Q^{(1)}_{ij}, Q^{(n)}_{kl} \}$ comes
exclusively from terms like
\beq
\int dx\left( \{Q_{ij}^{(1)},(j_{\mu})_{ka}\}2\dum S_{al}-(k
\!\!\leftrightarrow \!\!l)\right) \qquad .
\eeq

\vskip .5truecm
(c) Using the definition of $Q^{(1)}$ and the elementary current algebra and
dropping non-linear terms, we verify that

\newpage
\beq
\int \!\! dx\!\{Q_{ij}^{(1)},(j_{{}_0})_{ka}\}2\dum S_{al}\! -\! (k
\!\!\leftrightarrow \!\! l)
=\left( I\circ \!\int
\!\! dx[j_{{}_1}2\dum S+j_{{}_0}2\dum (j_{{}_0}2\dum S)
+2jS]\right) _{ij,kl} \quad .
\eeq
\[
\hskip 1cm \epsfbox{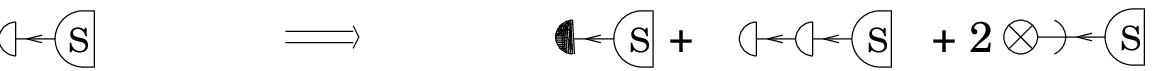}
\]

\beq
\int \!\! dx\! \{Q_{ij}^{(1)},(j_{{}_1})_{ka}\}2\dum S_{al}\! -\! (k
\!\! \leftrightarrow \!\! l)
=\left( I\circ \!\int \!\! dx[2j_{{}_0}j2\dum S+j_{{}_0}
2\dum (j_{{}_1}\dum S)]\right) _{ij,kl} \quad .
\eeq
\[
\hskip .7cm\epsfbox{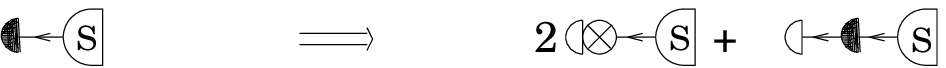}
\]

\noindent where some new symbols were introduced,
\[ j \Longleftrightarrow \;  \raise -.1cm \hbox{\epsfbox{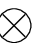}}
\]
\beq {1\over 2}\partial \Longleftrightarrow \;
 \raise -.1cm \hbox{\epsfbox{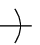}}
\hskip 1truecm \, \hskip 1truecm
\raise -.1cm \hbox{\epsfbox{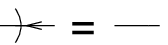}}\; \Longleftrightarrow
{1\over 2} \partial 2\dum = I
\eeq

\noindent
The previous expressions justify the following prescription:
\vskip .3truecm

i) We start from the diagram of $Q^{(n)}$.

\vskip .5truecm
ii) Then we replace the left ``tip" of each chain according to the rules:

\beq
\epsfbox{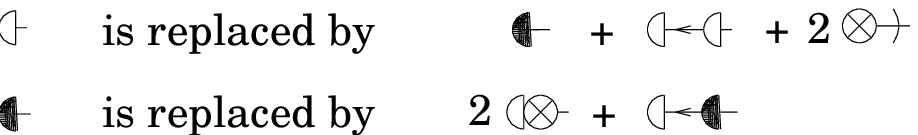}
\eeq

\vskip .5truecm
iii) The resulting diagram corresponds to $Q^{(n+1)}$.

\vskip .5truecm

\noindent
We remark that the substitution rules above
can be directly read from the following basic brackets:
\begin{eqnarray}
& &\{Q_{ij}^{(1)},(j_{{}_0})_{kl}\}=(I\circ
j_{{}_1}-2\dum j_{{}_0}j_{{}_0}-\partial j)_{ij,kl} +\cdots \\
& &\{Q_{ij}^{(1)},(j_{{}_1})_{kl}\}=(I\circ
j_{{}_0}j-2\dum j_{{}_0}j_{{}_1})_{ij,kl} +\cdots
\end{eqnarray}
In addition, one should not forget the constraints satisfied
by the $O(N)$ current $j_\mu $, given below,

\beq
[j_{\mu},j]_{+}=j_{\mu} \hskip 2truecm \raise -.1cm \hbox{\epsfbox{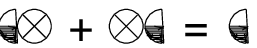}}
\eeq
\vskip 0.5cm
\beq
\int \, 2j_{{}_1}jS=\int \, (j_{{}_1}S+j\partial S) \hskip 2truecm
\raise -.1cm \hbox{\epsfbox{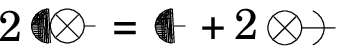}}
\eeq

\noindent
The half-white/half-black semicircle means $j_{{}_0}$ or $j_{{}_1}$
generically. We have tested the efficiency
of this method: comparing to the explicit algebraic algorithm in ref. [29]
we have taken much less time and space to construct the improved charges. For
the sake of clarity we have gathered a few examples in appendix A.

\vskip .3truecm
We have also developed a diagrammatic technique to calculate the algebra
itself. It can be seen as a set of contraction rules between the chains that
constitute the charges. Indeed,
in computing the algebra of non-local charges we have to consider all possible
``contractions" (i.e. Dirac brackets) between symmetrized chains. After some
partial integrations we end up with elementary contractions of the following
general kind:
\beq
\int \!\! \dum S_{ia}(x)\dum T_{bj}(x)
\{(j_{\mu})_{ab}(x),(j_{\nu})_{cd}(y)\}
\dum U_{kc}(y)\dum V_{dl}(y)\! -\!(i \!\!\leftrightarrow \!\! j)\! -\!
(k\!\! \leftrightarrow \!\!l) \quad .
\eeq
The current algebra (3) tells us that a contraction
$\{ j_\mu (x),j_\nu (y)\} $
may produce a current-like term $(I\circ j_\alpha )\delta (x-y)$ or a
Schwinger term. Let us discuss the first kind, in which case (28) produces 4
terms:
\begin{eqnarray}
& &\int dx \left[(\dum S\dum U^{{}^t}\circ \dum T^{{}^t}j_{\alpha}
\dum V)+(\dum T^{{}^t}\dum U^{{}^t}\circ
\dum Sj_{\alpha}\dum V) \right. + \nonumber \\
&+&\left. (\dum Sj_{\alpha}\dum U^{{}^t}\circ
\dum T^{{}^t}\dum V)+
(\dum T^{{}^t}j_{\alpha}\dum U^{{}^t}\circ
\dum S\dum V)\right]_{ij,kl} \quad .
\end{eqnarray}
We can associate each of the 4 terms above to one of the 4 possible
contractions between the 2 pairs of symmetrized chains. In the presence of a
Schwinger term we must take into account extra contributions involving the
intertwiner and partial integrations. In any event, the contractions between
chains can be resumed by the following rules:

\vskip .5truecm
{\bf Step 1: Choice}
\vskip .2truecm

In calculating $\{ Q^{(m)}, Q^{(n)}\} $ we
take one chain from $Q^{(m)}$ and other from $Q^{(n)}$.
Then we pick up the ``internal" current components we intend to
contract. This is symbolized by the generic diagram below:

\beq
\epsfbox{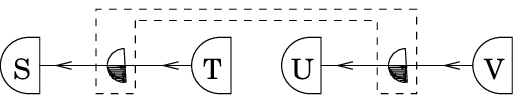}
\eeq

\vskip .5truecm
{\bf Step 2: Isolation}
\vskip .2truecm

In each chain we must ``localize" the current components
chosen in step 1. This was explicitly made in (28) by means of partial
integrations. Within the diagrams this
is achieved by inverting some arrows until all of them are pointing towards the
chosen semicircle (i.e. the current component we are isolating).
Eventually a minus sign will be picked up, depending on the number
of inversions. Finally have have this sort of diagrams:

\beq
\epsfbox{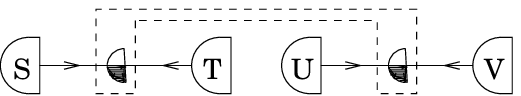}
\eeq

\newpage
{\bf Step 3: Bending}
\vskip .2truecm

The next step is just a graphic bending of chains, as a preparation to the
final contraction. The chains from (31) should be bended in the following way:

\beq
\epsfbox{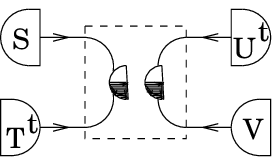}
\eeq

\noindent
Notice that the sub-chains $T$ and $U$ were transposed as eq. (31) demands.
Actually the graphic bending implies the transposition, as exemplified below

\beq
\epsfbox{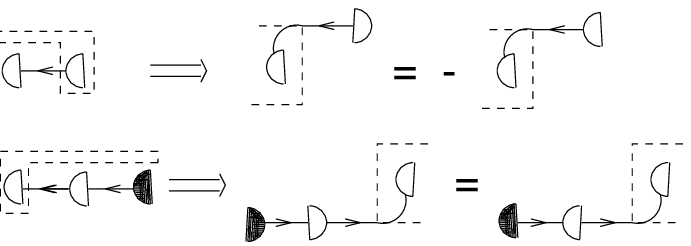}
\eeq

\noindent where the transposed current components are naturally represented as
\begin{eqnarray}
& & \0j ^{{}^t}= \raise -.1cm \hbox{\epsfbox{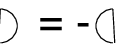}}= -\0j \nonumber \\
& & \1j ^{{}^t}= \raise -.1cm \hbox{\epsfbox{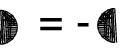}}= -\1j
\end{eqnarray}

\vskip .5truecm
{\bf Step 4: Contraction}
\vskip .2truecm

Finally we perform the contraction in (32) according to the rules below:

\beq
\epsfbox{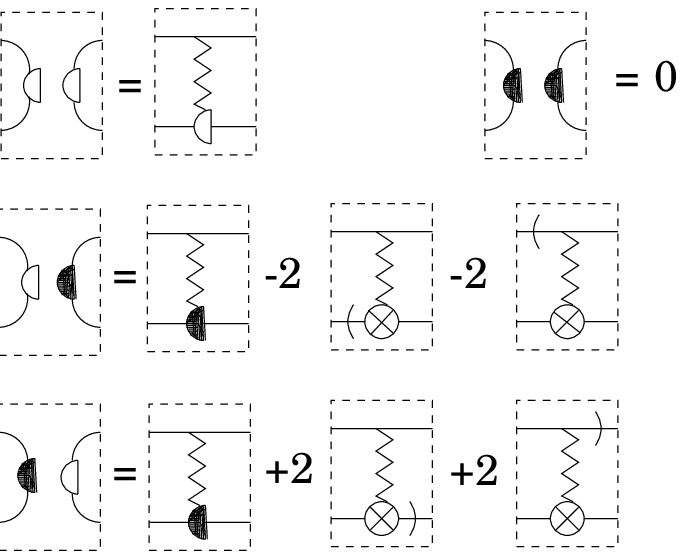}
\eeq
\noindent
where we introduced a symbol corresponding to the $\circ $-product,

\beq
\raise -.6cm \hbox{\epsfbox{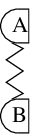}} =\int(A\circ B)=\int(B\circ A) \quad .
\eeq

\noindent
For instance, a typical contraction between $\0j $ components would be

\[
\epsfbox{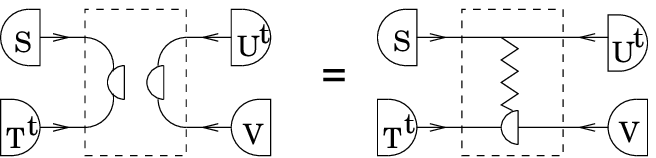}
\]
\beq
=\int dx(\dum S\dum U^{{}^t}\circ
\dum T^{{}^t}j_{{}_0}\dum V) \quad .
\eeq

\noindent
Of course one must repeat all steps for every possible contraction.

\vskip .2truecm
The current $j_\mu $ obeys another constraint [27] involving the
$\circ $-product, namely

\beq
(j_{\mu}\circ j)= 0 = \raise -.6cm \hbox{\epsfbox{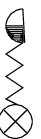}}
\eeq

\noindent
which must also be taken into account.

\vskip .3truecm
We mention that the elementary contractions in (35) are nothing but the graphic
representation of the current algebra (3), where the diagrams containing the
intertwiner field come from Schwinger terms followed by partial
integrations. These rules were applied to compute various brackets and in all
cases the algebra (6) was confirmed. One can also find an example in appendix
A.

The most remarkable outcome of this graphic procedure is that it poses an easy
and straightforward way to the supersymmetric extension -- and possibly other
generalizations.
\section{ Supersymmetric model}

The supersymmetric non-linear $O(N)$ sigma model is defined [30-32] by the
Lagrangean

\beq
{\cal L}_s=\frac{1}{2}\partial_{\mu}\phi_{i}\partial^{\mu}\phi_{i}+
\frac{i}{2}\overline{\psi}_{i}\partial\!\!\! /\psi_{i}+
\frac{1}{8}(\overline{\psi}_{i}\psi_{i})^{2}\quad ,
\eeq
where $\phi _i$ are scalars and $\psi _i$ are Majorana fermions satisfying the
constraints
\beq
\sum_{i=1}^{N}\phi_{i}^{2}-1=0 \quad , \quad \sum_{i=1}^{N}\phi_{i}\psi_{i}=0
 \quad .
\eeq
We also have a conserved $O(N)$ current $J_\mu $
which can be split into bosonic and fermionic parts
\begin{eqnarray}
& &J_{\mu}=j_{\mu}+b_{\mu} \quad ,\quad \partial _\mu J^\mu =0\quad ,
\nonumber \\
& &(j_{\mu})_{ij}=\phi_{i}\partial_{\mu}\phi_{j}-
\phi_{j}\partial_{\mu}\phi_{i} \quad ,\\
& &(b_{\mu})_{ij}=-i\overline{\psi}_{i}\gamma_{\mu}\psi_{j} \quad .\nonumber
\end{eqnarray}
whose curvature obeys the equation
\beq
F_{\mu \nu} = \partial _\mu J_\nu -\partial _\nu J_\mu +2[J_\mu ,J_\nu ]=
-(\partial _\mu b_\nu - \partial _\nu b_\mu ) \quad .
\eeq
Even though its curvature is not null, one can construct non-local conserved
charges out of $J_\mu$ [30,31]. Here we shall deal with an algebraic procedure
to derive these charges. Therefore it is necessary to start from the
elementary $O(N)$ current algebra, listed below,

\begin{eqnarray}
& &\{(j_{{}_0})_{ij}(x),(j_{{}_0})_{kl}(y)\}=[(I\circ j_{{}_0})-
(j\circ b_{{}_0})]_{ij,kl}(x)\delta(x-y) \quad ,\nonumber \\
& &\{(j_{{}_0})_{ij}(x),(j_{{}_1})_{kl}(y)\}=
(I\circ j_{{}_1})_{ij,kl}(x)\delta(x-y)+(I\circ j)_{ij,kl}(y)\delta'(x-y)
\quad ,\nonumber \\
& &\{(j_{{}_1})_{ij}(x),(j_{{}_1})_{kl}(y)\}=0 \quad ,\nonumber \\
& & \nonumber \\
& & \{(b_{{}_0})_{ij}(x),(b_{{}_0})_{kl}(y)\}=
[(I\circ b_{{}_0})-(j\circ b_{{}_0})]_{ij,kl}(x)\delta(x-y) \quad ,\nonumber \\
& &\{(b_{{}_0})_{ij}(x),(b_{{}_1})_{kl}(y)\}=
[(I\circ b_{{}_1})-(j\circ b_{{}_1})]_{ij,kl}(x)\delta(x-y) \quad ,\\
& &\{(b_{{}_1})_{ij}(x),(b_{{}_1})_{kl}(y)\}=
[(I\circ b_{{}_0})-(j\circ b_{{}_0})]_{ij,kl}(x)\delta(x-y) \quad ,\nonumber \\
& & \nonumber \\
& & \{(j_{{}_0})_{ij}(x),(b_{{}_0})_{kl}(y)\}=
(j\circ b_{{}_0})_{ij,kl}(x)\delta(x-y) \quad ,\nonumber \\
& &\{(j_{{}_0})_{ij}(x),(b_{{}_1})_{kl}(y)\}=
(j\circ b_{{}_1})_{ij,kl}(x)\delta(x-y) \quad ,\nonumber \\
& &\{(j_{{}_1})_{ij}(x),(b_{{}_0})_{kl}(y)\}=0 \quad ,\nonumber \\
& &\{(j_{{}_1})_{ij}(x),(b_{{}_1})_{kl}(y)\}=0 \quad ,\nonumber
\end{eqnarray}
where the intertwiner and the $\circ $-product were already
defined in (4) and (5).
The $O(N)$ local charge and the first non-local charge are given [30,31] by the
integrals
\begin{eqnarray}
& &Q^{(0)}=\int dx(j_{{}_0}+b_{{}_0}) \nonumber \\
& &Q^{(1)}=\int dx(j_{{}_1}+2b_{{}_1}+
2(j_{{}_0}+b_{{}_0})\dum (j_{{}_0}+b_{{}_0})) \quad .
\end{eqnarray}
Some other supersymmetric {\it standard} non-local charges can be found in the
literature [30-32]. However, as in the bosonic case, we are searching for {\it
improved} charges satisfying the simplest algebra. Using the algebraic method
proposed in ref. [29] we have computed the improved charges and their brackets
up to $n=3$, finding the {\it same cubic algebra} given by (6). Calculation is
hopelessly longer than in the bosonic theory, but we have been able to
develop some graphic rules which rather simplified our work. This diagrammatic
method is a direct extension of the one proposed for the bosonic theory. For
instance, one can show that the supersymmetric step-generator $Q^{(1)}$
satisfies the following algebraic relations:

\begin{eqnarray}
& &\{Q^{(1)}_{ij},(j_{{}_0})_{kl}\}=\left( I\circ j_{{}_1}
-2\dum j_{{}_0}j_{{}_0} -2\dum b_{{}_0}j_{{}_0}- \partial j\right) +\cdots \\
& &\{Q^{(1)}_{ij},(j_{{}_1})_{kl}\}=\left( I\circ 2j_{{}_0}j
-2\dum j_{{}_0}j_{{}_1} -2\dum b_{{}_0}j_{{}_1}\right) +\cdots \\
& &\{Q^{(1)}_{ij},(b_{{}_0})_{kl}\}=\left( I\circ 2b_{{}_1}
-2\dum j_{{}_0}b_{{}_0}) -2\dum b_{{}_0}b_{{}_0}\right) +\cdots \\
& &\{Q^{(1)}_{ij},(b_{{}_1})_{kl}\}=\left( I\circ 2b_{{}_0}
-2\dum j_{{}_0}b_{{}_1}) -2\dum b_{{}_0}b_{{}_1}\right) +\cdots
\end{eqnarray}
As in the bosonic model (recall eqs.(23-25)) these relations lead us to the
proper transformation rules for the construction of charges. One can use the
following iterative procedure:
\vskip .3truecm

i) We propose the symbolic notation
\[j_{{}_0}\Longleftrightarrow \;
\raise -.1cm \hbox{\epsfbox{nd1.eps}}  \quad ,\quad
j_{{}_1} \Longleftrightarrow \;\raise -.1cm \hbox{\epsfbox{nd2.eps}} \]

\beq
b_{{}_0} \Longleftrightarrow \;
\raise -.1cm \hbox{\epsfbox{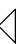}}  \quad ,\quad
b_{{}_1} \Longleftrightarrow \;\raise -.1cm \hbox{\epsfbox{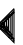}}
\eeq

\[2\dum  \Longleftrightarrow \;
\raise .06cm \hbox{\epsfbox{nd4.eps}}\quad ,\quad
{1\over 2}\partial \Longleftrightarrow \;
\raise -.1cm \hbox{\epsfbox{nd13.eps}}\]

\[j \Longleftrightarrow \;\raise -.1cm \hbox{\epsfbox{nd12.eps}} \]

\vskip .5truecm
ii) We take the diagram associated to $Q^{(n)}$ and replace the left ``tip" of
each chain as follows:

\beq
\epsfbox{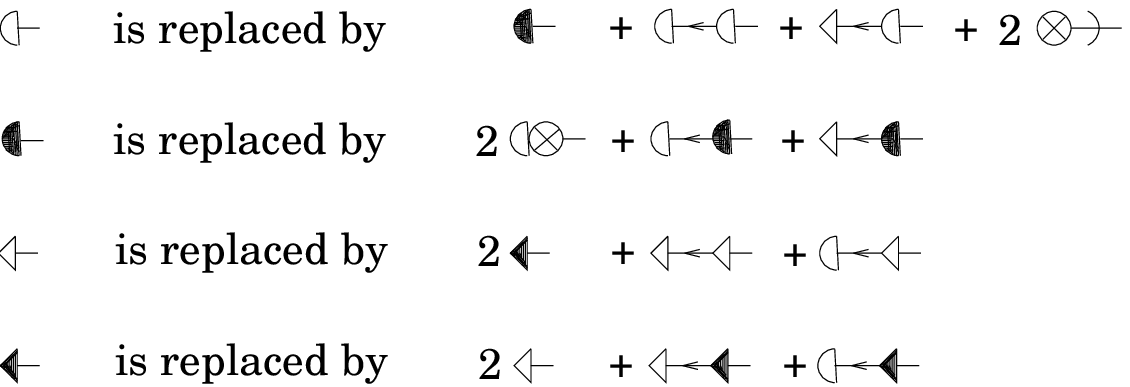}
\eeq

\noindent These transformations are a direct translation of eqs. (45-48).

\vskip .5truecm
iii) After using the constraints on $j_\mu $ and $b_\mu $ you will have the
diagram of $Q^{(n+1)}$.
\vskip .5truecm

In order to calculate the algebra
between the non-local charges, one should follow the
same algorithm (choice, isolation, bending and contraction), using the
contraction rules below


\[
\epsfbox{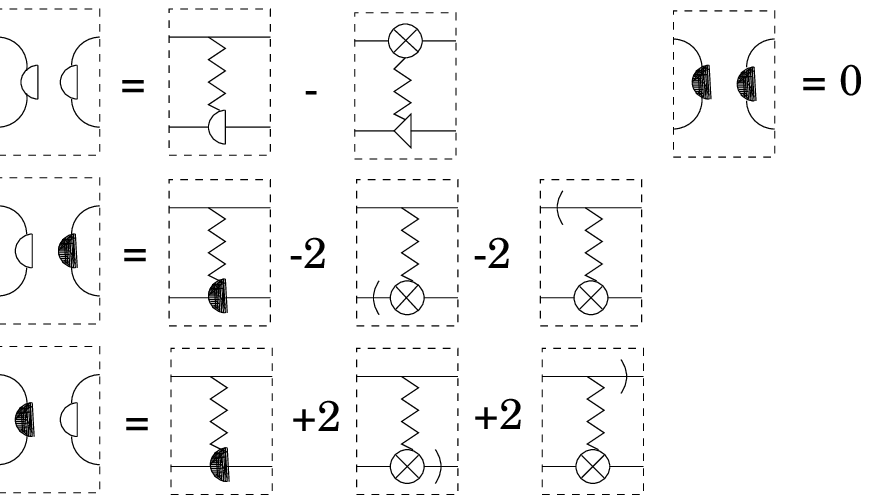}
\]

\[
\epsfbox{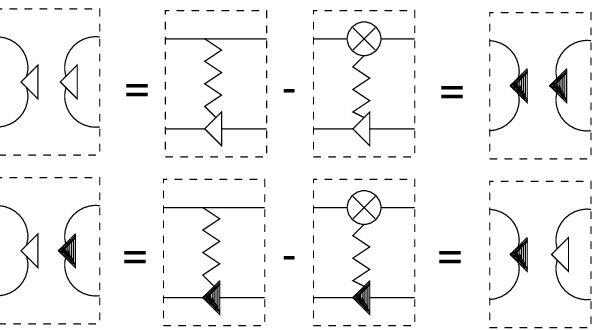}
\]

\beq
\epsfbox{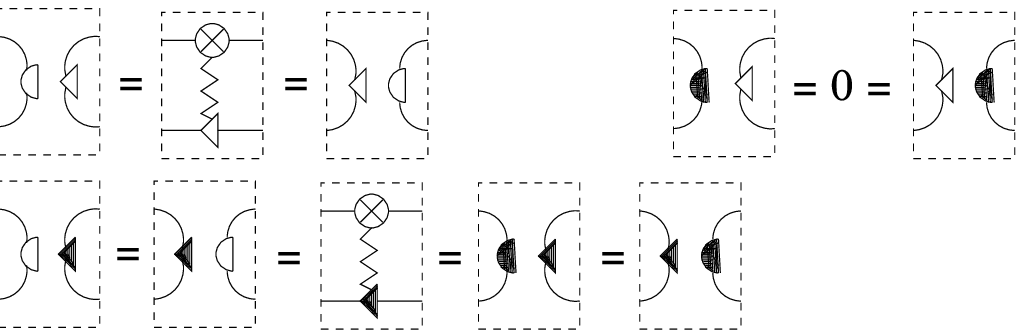}
\eeq

\noindent
which is the graphic version of the algebra (43). We also have a new
constraint,

\beq
b_{\mu}j=jb_{\mu}= 0 = \raise -.1cm \hbox{\epsfbox{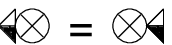}}
\eeq

\noindent to be added to the list in (26,27,38). As before, the
half-white/half-black triangle means $b_{{}_0}$ or $b_{{}_1}$ in general.

\vskip .5truecm
We have used this procedure to construct several charges and confirmed the
algebra (6). This is actually the main result of this paper, confirming
previous conjectures.

\vskip .3truecm
To complete the algebraic analysis, we have also considered the
conserved supersymmetry current and charge, given by
\begin{eqnarray}
& &{\cal J}_\mu = \partial \!\!\! / \phi _i \gamma _\mu \psi _i \quad ,
\nonumber \\
& &{\cal Q} = \int dx\, {\cal J}_{{}_0}\quad .
\end{eqnarray}
Using the equations of motion, we have checked that
\beq
\{ {\cal Q}, Q^{(n)} \} = 0 \quad , \quad n\ge 0
\eeq
which means that every non-local charge is invariant under supersymmetry
on-shell -- as already pointed in ref. [32].
Therefore the non-local charges in the
supersymmetric sigma model are all bosonic. However we must stress that this
is not a general property: for instance, in ref. [34] one finds an integrable
supersymmetric theory -- the supersymmetric two boson hierarchy -- containing
fermionic non-local charges whose graded algebra exhibits cubic terms similar
to those of eq. (6). It would be very interesting to develop graphic rules for
this kind of model [35].

\section{Improved charges in the $O(N)$ Gross-Neveu model}

This model consists of an $N$-plet of Majorana fermions transforming as a
fundamental representation of the $O(N)$ group, with a quartic interaction. Its
Lagrangean reads [7]
\beq
{\cal L}_{{}_{GN}} =
\frac{i}{2}\overline{\psi}_{i}\partial\!\!\! /\psi_{i}+
\frac{1}{8}(\overline{\psi}_{i}\psi_{i})^{2}\quad ,
\eeq
and it can be regarded as the limit of null bosonic field ($\phi _i \to 0$) in
the supersymmetric model (39) -- notice that the constraints (40) disappear.
The Noether current associated to the $O(N)$ rotations is
\beq
(b_\mu )_{ij} = -i\overline{\psi}_{i}\gamma_{\mu}\psi_{i} \quad ,\quad \partial
_\mu b^\mu =0
\eeq
and it satisfies the curvature-free condition
\beq
\partial _\mu b_\nu -\partial _\nu b_\mu + [b_\mu ,b_\nu ] = 0
\eeq
and the algebraic relations
\begin{eqnarray}
& & \{(b_{{}_0})_{ij}(x),(b_{{}_0})_{kl}(y)\}=
(I\circ b_{{}_0})_{ij,kl}(x)\delta(x-y) \nonumber \\
& &\{(b_{{}_0})_{ij}(x),(b_{{}_1})_{kl}(y)\}=
(I\circ b_{{}_1})_{ij,kl}(x)\delta(x-y) \\
& &\{(b_{{}_1})_{ij}(x),(b_{{}_1})_{kl}(y)\}=
(I\circ b_{{}_0})_{ij,kl}(x)\delta(x-y) \nonumber
\end{eqnarray}
As before, we may construct an infinite number of conserved non-local currents
using the potential algorithm: we
consider a conserved current $B_\mu ^{(n)}$ and
the corresponding potential $\xi ^{(n)}$,
\beq
B_\mu ^{(n)}=\epsilon _{\mu \nu } \partial ^\nu \xi ^{(n)}\quad .
\eeq
Then we define the current $B_\mu ^{(n+1)}$ as
\beq
B_\mu ^{(n+1)} = 2( \partial _\mu + b_\mu ) \xi ^{(n)}\quad .
\eeq
The properties (56) and (57) imply that $B_\mu ^{(n+1)}$ is also conserved.
Starting out with $B_\mu ^{(0)} = b_\mu $ we find an infinite number of
conserved charges $Q^{(n)}=\int dx\, B_o^{(n)}$. After applying this
algorithm to build up some of them, it is straightforward to check
that this method is equivalent to the following graphic procedure:
one chooses some symbols to represent $b_{{}_0}$ and $b_{{}_1}$, for instance,
\beq
b_{{}_0}  \Longleftrightarrow \;
\raise -.1cm \hbox{\epsfbox{nd28.eps}} \hskip 2truecm
b_{{}_1}  \Longleftrightarrow \;\raise -.1cm \hbox{\epsfbox{nd29.eps}}
\eeq
then one takes the sequence of chains associated to $Q^{(n)}$ and uses the
replacement rules for left-tips

\beq
\epsfbox{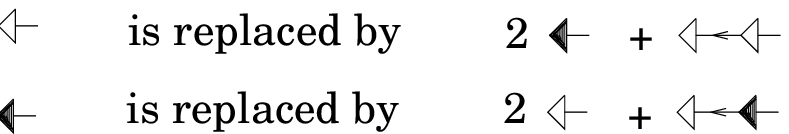}
\eeq

\noindent
On the other hand, this is precisely the limit $\phi _i\to 0$ of the
transformation rules (50) in the supersymmetric theory.

\vskip .3truecm
This provides an alternative derivation of the graphic rules to construct
charges in the Gross-Neveu model. Moreover it implies that those charges
defined by the algorithm (60) are actually the improved charges and thus
they must obey the cubic algebra (6).

\newpage

\section{ Conclusions and final remarks}

Diagrammatic methods are frequently used in physics to simplify long
calculations and so we have proposed a graphic procedure to construct and
compute the algebra of non-local charges in non-linear sigma models.
Applying such procedure we have been able to verify that
the (improved) non-local charges in the supersymmetric $O(N)$ sigma model obey
a cubic Yangian-like algebra which can be expressed as in eq. (8).

\vskip .2truecm
One could easily recover the bosonic model from the supersymmetric theory by
taking the no-fermion limit $\psi _i \to 0$. Moreover,
the $O(N)$ invariant Gross-Neveu model
can be obtained after erasing the bosonic fields ($\phi _i \to 0$).
The improved charges in these models may be different but
their algebra is exactly the same. In the Gross-Neveu model, the improved
charges could be computed by means of two different methods and therefore the
corresponding graphic rules could be confirmed and further understood.
We expect to find a similar confirmation in the sigma models but the presence
of the intertwiner field within the diagrams has impeded us so far.

\vskip .2truecm
It is also interesting to consider
the inclusion of Wess-Zumino terms, which modify the
current algebra (see for instance ref. [28,29]) and derive the corresponding
graphic rules. This problem and the general application of diagrammatic methods
to integrable theories is presently under investigation [35].

\vskip 2truecm
\noindent {\bf Acknowledgements}
\vskip .5truecm

\noindent
We would like to thank E. Abdalla and M.C.B. Abdalla for helpful comments. We
also thank J.C. Brunelli for suggestions and for his participation in the early
stages of this work.

\newpage

\appendix

\section{Diagrammatic rules: examples}

\subsection{Improved non-local charges $Q^{(n)}$ in the non-linear
sigma model}

Let us apply the graphic method to construct some of the charges in the bosonic
model:

\vskip .2truecm
i) $n=0$
\vskip .2truecm

According to our conventions the $O(N)$ local charge is represented as

\beq
Q^{(0)}= \int dx\, \0j \Longleftrightarrow \; \raise -.1cm \hbox{
\epsfbox{nd1.eps}}
\eeq

\vskip .2truecm
ii) $n=1$
\vskip .2truecm

Using the transformation rules (23) we obtain

\beq
\epsfbox{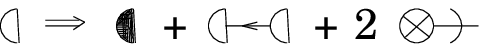}
\eeq

\noindent
The last term is zero, since it should be read as $\int dx\, j\partial I = 0$.
Therefore we have the following diagram for $Q^{(1)}$:

\beq
\epsfbox{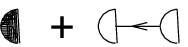}
\eeq

\noindent which means that the first non-local charge is written as
\beq
Q^{(1)}=\int dx\, (\1j +2\0j \dum \0j )
\eeq

\vskip .2truecm
iii) $n=2$
\vskip .2truecm

Now we take the first chain from $Q^{(1)}$ and apply the replacement rule
(23) again,

\beq
\epsfbox{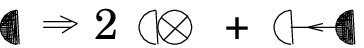}
\eeq

\noindent Next we transform the second chain, replacing its left tip according
to (23)

\beq
\epsfbox{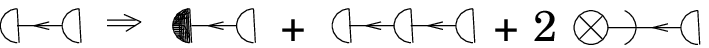}
\eeq

\noindent Recalling the property (22),

\beq
\epsfbox{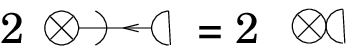}
\eeq
adding all contributions and remembering the constraint

\beq
\epsfbox{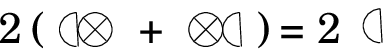}
\eeq

\noindent the resulting diagram is

\beq
\epsfbox{nd9.eps}
\eeq

\noindent and therefore the second non-local charges reads

\beq
Q^{(2)}=\int dx\left[ 2\0j + 2\0j \dum \1j  + 2\1j \dum \0j  + 4\0j \dum
(\0j \dum \0j ) \right]
\eeq

\noindent Notice that the proper use of the constraints has rendered a charge
free of intertwiners. This property has been checked up to $n=7$ and we expect
it to hold for every $n$.

\subsection{ Graphic derivation of $\{ Q^{(1)},Q^{(1)}\} $ in the non-linear
sigma model}

In this case we must consider all possible contractions between the sequence of
chains from $Q^{(1)}$,

\beq
\epsfbox{nd35.eps}
\eeq

\noindent which altogether sums 9 contractions. Here they are:

\vskip .2truecm
i)

\beq
\epsfbox{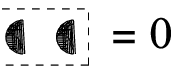}
\eeq

\vskip .2truecm
ii)

\beq
\epsfbox{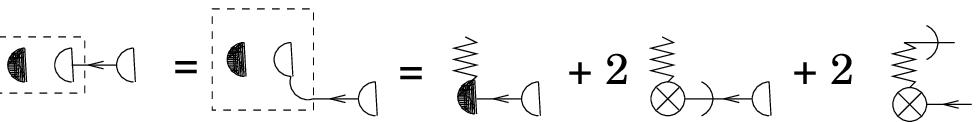}
\eeq

\noindent The first term corresponds to

\beq
\raise -.4cm \hbox{\epsfbox{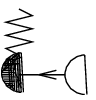}}= \int (I\circ 2\1j \dum \0j )
\eeq

\noindent and the second one reads

\beq
\raise -.3cm \hbox{\epsfbox{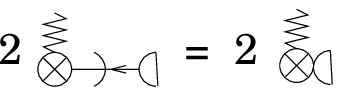}}= \int (I\circ 2j\0j )
\eeq

\noindent The last term vanishes because it contains a derivative of the
identity matrix,

\beq
\raise -.4cm \hbox{\epsfbox{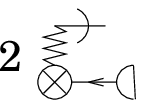}} = 2\int (\partial I\circ j\dum \0j ) =0
\eeq

\vskip .2truecm
iii)

\beq
\epsfbox{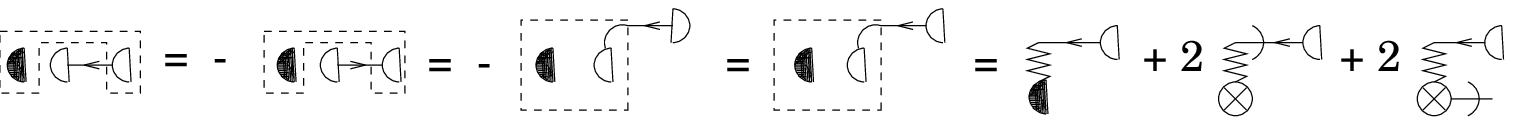}
\eeq

\noindent Notice that we have computed two minus signs: one from the inversion
of an arrow (along the isolation step) and other from the transposition of $\0j
$ (during the bending step). The first contribution is

\beq
\raise -.5cm \hbox{\epsfbox{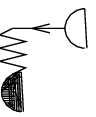}} = \int (2\dum \0j \circ \1j )
\eeq

\noindent The second one vanishes due to the constraint (38)

\beq
\epsfbox{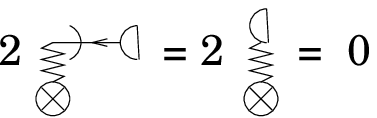}
\eeq

\noindent and the third term is null too,

\beq
\raise -.5cm \hbox{\epsfbox{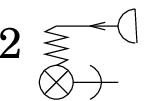}}= \int dx (2\dum \0j \circ j\partial I)
= 0
\eeq

\noindent Let us proceed with the remaining contractions:

\vskip .2truecm
iv)

\beq
\epsfbox{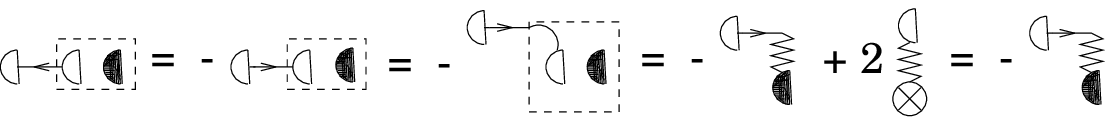}
\eeq

\vskip .2truecm
v)

\beq
\epsfbox{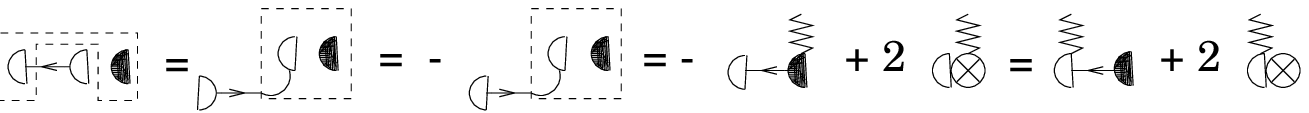}
\eeq

\vskip .2truecm
vi)

\beq
\epsfbox{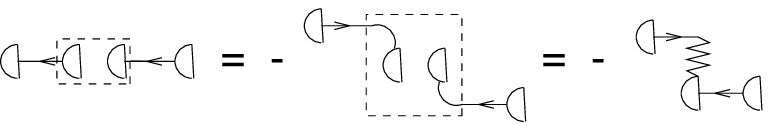}
\eeq

\vskip .2truecm
vii)

\beq
\epsfbox{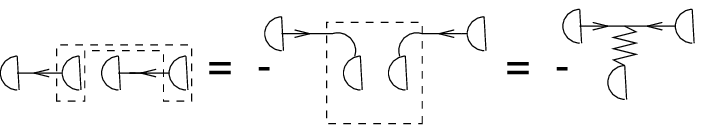}
\eeq

\vskip .2truecm
viii)

\beq
\epsfbox{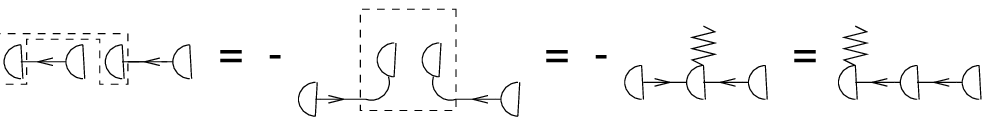}
\eeq

ix)

\beq
\epsfbox{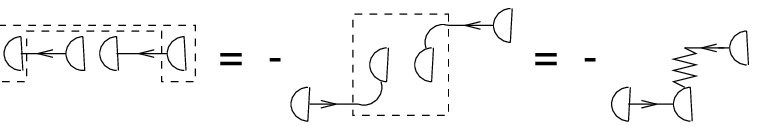}
\eeq

\newpage
\noindent Adding all the non-vanishing contributions we end up with the
following terms:

\beq
\epsfbox{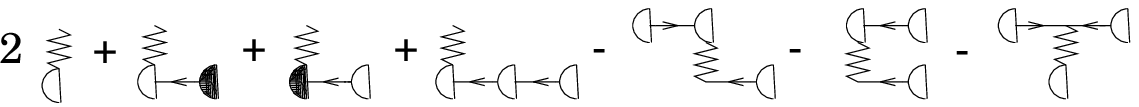}
\eeq

\noindent The linear part of the algebra is clearly recognized

\begin{eqnarray}
& &\hskip 2.05cm \epsfbox{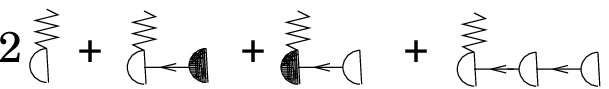} \nonumber\\
& &=\int dx\, (I\circ 2\0j + 2\0j \dum \1j + 2 \1j \dum \0j + 4 \0j \dum (\0j
\dum \0j ))\nonumber \\
& &=(I\circ Q^{(2)})
\end{eqnarray}

\noindent The remaining three diagrams provide the surface term that
corresponds to the cubic part of the algebra:

\beq
\epsfbox{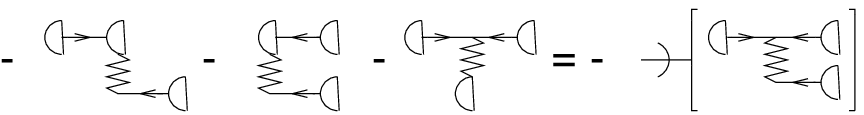}
\eeq

\[
= - \int dx\, {1\over 2}\partial \, 8\left( \dum \0j \dum \0j \circ \dum \0j
\right) = - (Q^{(0)} Q^{(0)} \circ Q^{(0)})
\]

\noindent Finally we obtain the expected answer,
\beq
\{ Q^{(1)}, Q^{(1)}\} = (I\circ Q^{(2)}) - (Q^{(0)} Q^{(0)} \circ Q^{(0)})
\eeq

This commented example may seem rather long, not revealing the
actual power of the graphic method. But we can assure the reader that, after
practicing the contraction rules for a while -- and skipping the intermediate
comments -- we have been able to compute many Dirac brackets very efficiently.


\begin{thebibliography}{99}

\bibitem{1}
L\"uscher, M., Pohlmeyer, K.: Scattering of massless lumps and
non-local charges in the two-dimensional classical non-linear $\sigma$-model.
Nucl. Phys. {\bf B137}, 46-54 (1978).

\bibitem{2}
L\"uscher, M.: Quantum non-local charges and absence of particle
production in two-dimensional non-linear $\sigma$-model. Nucl. Phys.
{\bf B135}, 1-19 (1978).

\bibitem{3}
de Vega, H.J.: Field theories with an infinite number of conservation
laws and\break B\"acklund transformations in two dimensions.
Phys. Lett. {\bf 87B}, 233-236 (1979).

\bibitem{4}
Pohlmeyer, K.: Integrable Hamiltonian systems and interactions
through quadratic constraints. Commun. Math. Phys. {\bf 46}, 207-221 (1976).

\bibitem{5}
Br\'ezin, E., Itzykson, C., Zinn-Justin, J., Zuber, J.B.: Remarks
about the existence of non-local charges in two-dimensional models. Phys. Lett.
{\bf 82B}, 442-444 (1979).

\bibitem{6}
Buchholtz, D., Lopuzanski, J.T.: Non-local charges: a new concept in
quantum field theory. Lett. Math. Phys. {\bf 3}, 175-180 (1979).

\bibitem{7}
Abdalla, E., Abdalla, M.C.B., Rothe, K.: Non-perturbative
methods  in two-dimen\-sional quantum field theory. Singapore: World Scientific
1991.

\bibitem{8}
Zamolodchikov, A.B.,  Zamolodchikov, Al.B.: Factorized S-Matrices in
two dimensions as the exact solutions of certain relativistic quantum field
theory models. Ann. Phys. {\bf 120}, 253-291 (1979).

\bibitem{9}
Mussardo, G.: Off-critical statistical models: factorized scattering
theories and bootstrap program. Phys. Rep. {\bf 218}, 215-379 (1992).

\bibitem{10}
Abdalla, E., Abdalla, M.C.B., Sotkov, G., Stanishkov, M.: Off
critical current algebras. Int. J. Mod. Phys. {\bf A10}, 1717-1736 (1995).

\bibitem{11}
Abdalla, E., Forger, M., Gomes, M.: On the origin of anomalies in
the quantum non-local charge for the generalized non-linear sigma models. Nucl.
Phys. {\bf B210}, 181-192 (1982).

\bibitem{12}
Dolan, L.: Kac-Moody algebra is hidden symmetry of chiral models.
Phys. Rev. Lett. {\bf 47}, 1371-1374 (1981).

\bibitem{13}
Gomes, M. and Ha, Y.K.: Remarks on the algebra for higher non-local
charges. Phys. Rev. {\bf D28}, 2683-2685 (1983).

\bibitem{14}
de Vega, H.J., Eichenherr, H., Maillet, J.M.: Classical and quantum
algebras of non-local charges in $\sigma$-models. Commun. Math. Phys. {\bf 92},
507-524 (1984).

\bibitem{15}
de Vega, H., Eichenherr, H., Maillet, J.M.: Yang-Baxter
algebras of monodromy matrices in integrable quantum field theories. Nucl.
Phys. {\bf B240}, 377-399 (1984).

\bibitem{16}
Maillet, J.-M.: Hamiltonian structures for integrable classical
theories from graded Kac-Moody algebras. Phys. Lett. {\bf 167B}, 401-405
(1986);
New integrable canonical structures in two-dimensional models. Nucl. Phys.
{\bf B269}, 54-76 (1986).

\bibitem{17}
Drinfel'd, V.G.: Hopf algebras and the quantum Yang-Baxter equation.
Sov. Math. Dokl. {\bf 32}, 254-258 (1985).

\bibitem{18}
Drinfel'd, V.G.: A new realization of Yangians and quantized affine
algebras. Sov. Math. Dokl. {\bf 36}, 212-216 (1988).

\bibitem{19}
Bernard, D.: Hidden Yangians in 2-D massive current algebras.
Commun. Math. Phys. {\bf 137}, 191-208 (1991).

\bibitem{20}
Leclair, A., Smirnov, F.A.: Infinite quantum group symmetry of
fields in massive 2D quantum field theory. Int. J. Mod. Phys. {\bf A7},
2997-3022 (1992).

\bibitem{21}
Haldane, F.D., Ha, Z.N.C., Talstra, J.C., Bernard, D., Pasquier, V.:
Yangian symmetry of integrable quantum chains with long-range interactions and
a new description of states in conformal field theory. Phys. Rev. Lett. {\bf
69}, 2021-2025 (1992).

\bibitem{22}
Mackay, N.J.: On the classical origins of Yangian symmetry in
integrable field theory. Phys. Lett. {\bf B281}, 90-97 (1992),
erratum-ibid.{\bf B308}, 444 (1993).

\bibitem{23}
Mackay, N.J.: On the bootstrap structure of Yangian-invariant
factorized $S$-matrices. PRINT-92-0535 (DURHAM). Bulletin board:
hep-th - 9211091. In Tianjin 1992, Proceedings, Differential geometric
methods in theoretical physics 360-363.

\bibitem{24}
Bernard, D.: An introduction to Yangian symmetries. Int. J. Mod. Phys. {\bf
B7}, 3517-3530 (1993).

\bibitem{25}
Schoutens, K.: Yangian symmetry in conformal field theory. Phys. Lett. {\bf
B331}, 335-341 (1994).

\bibitem{26}
Basu-Mallick, B., Radamevi, P.: Construction of Yangian algebra through a
multideformation parameter dependent R matrix. IMSC-94-38 preprint, hep-th
9406194.

\bibitem{27}
Forger, M., Laartz, J., Sch\"aper, U.: Current algebra of classical
non-linear sigma models. Commun. Math. Phys. {\bf 146}, 397-402 (1992).

\bibitem{28}
Abdalla, E., Forger, M.: Current algebra of WZNW models at and away
from criticality. Mod. Phys. Lett. {\bf 7A}, 2437-2447 (1992).

\bibitem{29}
Abdalla, E., Abdalla, M.C.B., Brunelli, J.C., Zadra, A.: The Algebra of
Non-local Charges in Non-linear Sigma Models. Commun. Math.
Phys. {\bf 166}, 379-396 (1994).

\bibitem{30}
Curtright, T.L., Zachos, C.K.: Non-local currents for supersymmetric
nonlinear models. Phys. Rev. {\bf D21}, 411-417 (1980).

\bibitem{31}
Barcelos-Neto, J., Das, A., Maharana, J.: Algebra of charges in the
supersymmetric non-linear $\sigma$-model. Z. Phys. {\bf 30C}, 401-405 (1986).

\bibitem{32}
Curtright, T.L., Zachos, C.K.: Supersymmetry and the nonlocal yangian
deformation symmetry. Nucl. Phys. {\bf B402}, 604-612 (1993).

\bibitem{33}
Abdalla, E., Abdalla, M.C.B., Branco, O.H.G., Saltini, L.E.: Current
Algebra of Super WZNW Models. J. Phys. A, {\bf 27}, 4709-4715 (1994).

\bibitem{34}
Brunelli, J.C., Das, A.: Properties of nonlocal charges in the supersymmetric
two boson hierarchy. Phys. Lett. {\bf B354}, 307-314 (1995).

\bibitem{35}
Saltini, L.E., Zadra, A.: work in progress.

\end{thebibliography}
\end{document}